# *Émergence et fragilité d'une recherche-création (2000-2007)*

Par **Georges Gagneré**, in *Ligeia*, n°137-140, p. 148-158, https://doi.org/10.3917/lige.137.0148

Ma démarche de recherche-création coïncide avec la rencontre du « paradigme numérique » et la tentative de son incorporation au fondement de mon écriture scénique. J'ai abordé le numérique sur scène après mes trente ans, au moment où les ordinateurs grand public devenaient assez puissants pour calculer les images et les sons en temps réel et s'immiscer dans l'ici et maintenant du jeu des acteurs, à l'aube des années 2000[1]. Je propose ici, du point de vue du metteur en scène, de rendre compte de l'évolution de la manière d'écrire scéniquement mes spectacles entre 2000 et 2007, en spécifiant les recherches qui l'ont influencée. Je décrirai ma participation au lancement d'un chantier de recherche d'ambition nationale encore en cours et je formulerai quelques remarques sur les fragilités qui peuvent surgir dans un « laboratoire technologique de mise en scène ».

## L'irruption du numérique dans une écriture scénique (2000-2001)

Ma rencontre avec l'univers numérique est liée à l'utilisation de l'image sur le plateau théâtral. Il s'agissait d'un travail sur les transformations économiques et politiques liées aux évolutions du capitalisme et notamment à l'impact des nouvelles technologies (quelques mois après la création du spectacle, l'éclatement de la « bulle Internet » déclenchait plusieurs années de crise mondiale). Le spectacle, *H Manifeste[s]*[2], construit en miroir d'un site internet, mettait en scène la dissolution du personnage de Madame H dans la société d'hypercommunication. Pendant un monologue, trois caméras fixes filmaient l'actrice du plateau vers le plateau (mini caméra située sur une télévision constituant un élément du décor, cf. figure 1 a), du plateau vers la salle (caméra située au-dessus de l'écran de rétroprojection, cf. figure 1 b), de la salle vers le plateau. Le flux des images en direct de Madame H était mixé à des éléments préenregistrés et diffusé simultanément sur la télévision et l'écran de rétroprojection, par une régie « analogique ». Les répétitions et les représentations avaient nécessité une très grande dextérité du régisseur et beaucoup de travail de production des images enregistrées en relation avec le jeu des comédiens. Les enjeux d'interaction étaient circonscrits au jeu de Madame H avec son image en direct, et à l'intervention du régisseur sur la conduite en fonction des actions scéniques.

Dans le spectacle suivant, *Huntsville, l'Ordre du Monde*[3], j'ai approfondi l'imbrication des images dans la scénographie avec une démultiplication des captations en direct des deux personnages, un gardien de « couloir de la mort » et la Femme en Bleu, allégorie de la mort. Deux caméras latérales fixes permettaient de filmer le décor dans sa profondeur, une caméra infrarouge (IR) était fixée sur pied au milieu du plateau, et une caméra dôme de vidéosurveillance, motorisée, télécommandable, fixée au gril, couvrait tout le décor. Un seul vidéoprojecteur illuminait de face l'élément principal du décor. Le dispositif de régie vidéo avait été décidé en concertation avec la créatrice vidéo suite à un workshop qu'elle avait effectué au mois d'août 2001 avec l'artiste numérique Pedro Soler, autour du logiciel expérimental NATO, réalisé à partir du logiciel Max[4], et permettant de transformer en temps réel des flux vidéo directs ou préenregistrés, d'une manière similaire à ce que Max-MSP permettait de faire sur les flux audio. La découverte de ce logiciel m'avait alors profondément impressionné car il laissait entrevoir de grandes possibilités d'écriture des images dans l'espace scénique en relation avec les actions des comédiens, et réalisables avec un ordinateur personnel de puissance standard. Non prévue initialement, la réalisation du spectacle avait nécessité la présence de Pedro Soler, pour assister la créatrice vidéo et construire une régie numérique *ad hoc* à partir d'un mélangeur Panasonic MX50 permettant de mixer des images préenregistrées et le flux direct de caméras modifié par un ordinateur Mac G4 en temps réel avec NATO.

---

[1] Il y a plus de dix ans, Louis-José Lestocart dressait déjà un panorama de ces bouleversement, de la « crise des fondements » des débuts du 20ème siècle aux différentes phases de la cybernétique, en passant par les mathématiciens de la logique formelle et Turing. Cf. Louis-José Lestocart, « Émergence – L'art en espace partagé » in *Ligeia, dossiers sur l'art*, « Art et Multimédia », N° 45-46-47-48, juillet-décembre 2003, Paris, pp. 152-155.

[2] Création du 18 au 21 mai 2000 aux XIèmes Rencontres Internationales de Théâtre de Dijon. Cf. www.didascalie.net/hmanifestes

[3] Création au Théâtre Gérard Philipe de Saint-Denis (salle Serreau) du 26 octobre au 11 novembre 2001. Cf. www.didascalie.net/huntsville

[4] Max est un logiciel graphique de programmation multimedia temps réel. Cf. http://cycling74.com/

Une séquence du spectacle illustre les possibilités d'imbrication des images, qui impliquent un jeu particulier des comédiens. Le gardien, enfermé dans son poste de surveillance, situé derrière un tulle, ressasse la description de la ville d'Huntsville, et à chaque cycle, une vignette avec son image en direct, captée par la caméra dôme et sur laquelle est appliqué un délai de quelques secondes, s'ajoute pour composer au final cinq flux vidéos en cascade, grâce au logiciel NATO. Cette image traduisant le ressassement mental du gardien est mixée à une captation en direct par la caméra IR de la Femme en Bleu, située physiquement au premier plan, qui surgit dans l'esprit du gardien et vient lui parler. Les deux acteurs, dans deux espaces différents, dialoguent alors par image interposée (cf. figure 1 c et d). De manière similaire au précédant spectacle, les acteurs étaient sollicités pour tourner des séquences en amont, rediffusées ultérieurement pendant la représentation, et à plusieurs moments ils devaient interagir avec leur image en direct. Cependant, l'utilisation du logiciel temps réel complexifiait l'écriture vidéoscénographique et imposait un positionnement spécifique des comédiens. En imbriquant une projection de leur jeu en direct dans l'espace, je sollicitais une amplification de leurs états émotionnels sur lesquels ils devaient cependant garder un contrôle. C'était une situation très contraignante, mais qui permettait de décaler l'interprétation vers une matérialisation spécifique de l'insoutenable tension d'un bourreau oeuvrant pour le « bien » de la société. Tout en étant expérimental et fragile, le dispositif vidéo a permis de jouer la série des représentations sans problème. En revanche, la construction des imbrications complexes d'images en direct au fil des répétitions a été plus laborieuse, et a conduit à une réflexion parallèle sur le dialogue entre mise en scène, création des images et régie numérique.

## Confrontation à l'outil informatique (2002-2003)

Dès la mise en chantier du spectacle suivant, *La Pensée*[5], Pedro Soler et moi-même avons proposé à un développeur, spécialiste de Max, de synthétiser une réponse logicielle à la méthodologie d'écriture vidéoscénographique que nous avions imaginée avec le régisseur pour améliorer le processus des répétitions et formaliser notre dialogue sur les flux vidéo. Le résultat, Mirage[6], permettait de sélectionner huit sources (des fichiers vidéos, des images, et jusqu'à deux caméras, selon les possibilités des cartes d'acquisition vidéo des ordinateurs,) auxquelles étaient appliqués des effets, puis un compositing sur douze couches (cf. figure 2 d). Chaque paramètre des modules de sources, d'effets et de compositing possédait une adresse spécifique qui pouvait être contrôlée de l'extérieur selon le protocole Open Sound Control (OSC) par un patch Max ou Pure Data (l'équivalent de Max en logiciel libre). Ce « moteur vidéo » et ses adresses constituaient ainsi le référentiel commun entre le metteur en scène, le créateur vidéo et le régisseur pour dynamiser les images en fonction de la scénographie et du jeu des acteurs.

En effet, la formalisation d'un accès temps réel aux paramètres visuels et la commercialisation émergente de hardware à bas coût permettant de porter des capteurs physiques sans fil, rendaient possible un contrôle direct des images par les comédiens. C'est ainsi qu'en accord avec la dramaturgie de *La Pensée*, les personnages interagissaient directement avec l'environnement vidéoscénographique. Christophe Caustier, jouant le personnage principal du docteur Kertjensev, portait trois capteurs : un bouton dans la main gauche, un flexomètre permettant de mesurer la fermeture du coude du bras gauche, et un accéléromètre permettant de mesurer l'orientation du poignet de la main droite selon deux directions. Au milieu du spectacle, Kertjensev se livrait à un chantage affectif envers Tatiana, la femme de son ami, dont il était passionnément amoureux. Il la piégeait dans la situation d'assister impuissante au meurtre de son mari. Cet enfermement était littéralement réalisé par une transformation progressive du décor (cf. figure 2 a), avec une caméra à l'arrière-plan filmant Tatiana enfermée, qui continuait le dialogue avec Kertjensev par l'intermédiaire de son visage en gros plan projeté sur la cage. A un moment précis, le régisseur enregistrait une séquence du flux direct, et Kertjensev, excité par son stratagème dément, manipulait cette séquence par des gestes de son bras droit (cf. figure 2 b). De *H Manifeste[s]* à *La Pensée*, l'implication du comédien dans l'écriture scénique des médias avait beaucoup progressé.

---

5  Création du 7 au 22 mars 2003 au Théâtre National de Strasbourg. Cf. www.didascalie.net/lapensee
6  Mirage était un standalone Max-SoftVNS, utilisant la librairie créée par l'artiste canadien David Rockeby, particulièrement adaptée pour le traitement des images en temps réel. Cf. www.didascalie.net/Mirage

Quatre caméras, dont deux motorisées, permettaient de filmer les acteurs sous différents angles. Trois vidéoprojecteurs, dont deux mobiles et manipulés par un comédien, permettaient de projeter des images de face et en rétroprojection sur les deux écrans constituant le décor. Ces deux écrans articulés étaient montés sur deux mats motorisés, eux même fixés sur une tournette pouvant effectuer un demi-tour. Les trois moteurs pilotant le décor, ainsi que trois caméras, étaient pilotés par un patch Max. Ainsi, images, sons, machinerie et capteurs étaient contrôlés par une régie homogène entièrement numérique (cf. figure 2 c). Le spectacle était techniquement très ambitieux, et les acteurs ont difficilement vécu les contraintes imposées, tout en réussissant au fil des répétitions à se les approprier et à en faire des moteurs de leur interprétation. La possibilité d'écrire les médias visuels et sonores, de contrôler les capteurs, les caméras, les vidéoprojecteurs et la machinerie avec le même environnement numérique, induit une complexité globale d'organisation et de synchronisation des actions, mais elle ouvre surtout des pistes dramaturgiques stimulantes.

En trois années, l'outil principal de l'écriture scénique que je déployais avec mes collaborateurs était devenu l'ordinateur, pourtant toujours considéré par la profession comme antinomique de la pratique théâtrale. En 2003, le spectacle vivant traversait une crise profonde liée à la modification du régime d'assurance chômage des intermittents du spectacle et les festivals de l'été avaient été annulés en signe de protestation. Ces mutations numérique et économique cristallisèrent au fil de l'année 2004 mes réflexions sur les enjeux d'écriture scénique émergents et sur la nécessité de mutualiser les moyens de recherche[7]. Ce désir utopique de mutualisation se rattachait à deux intuitions : celle de la mise en réseau d'une intelligence collective, très en vogue à l'époque, à partir des moyens offerts par le numérique et internet, afin d'élaborer des réponses collectives et génériques à des questionnements individuels et spécifiques ; et celle de rassembler des énergies et des ressources économiques afin de conduire les développements numériques nécessaires.

## Mise en réseau des expérimentations - didascalie.net (2004)

Dans la foulée du chantier de développement ouvert avec *La Pensée* par la création de Mirage, j'ai élargi ma réflexion sur la notion de logiciel moteur constituant un espace d'écriture des médias partagé par le metteur en scène, l'artiste numérique et le régisseur. Cette démarche a pris deux dimensions : d'un côté, la poursuite du développement de Mirage au niveau de la stabilité et des fonctionnalités d'utilisation, et de l'autre, la rencontre d'équipes artistiques à travers des étapes de co-développement et de co-création, que j'appelais *Escales*. L'organisation d'une semaine de rencontres internationales et de présentations d'environnements de création numérique, à l'occasion d'*Escale#4*[8], avait confirmé que les questionnements de mes collaborateurs et moi-même étaient en fait partagés par de nombreux artistes et structures. Le rapport à l'ordinateur et aux environnements logiciels pour écrire les médias et les interactions bouleversaient les pratiques.

Le besoin de développer les outils de création sur des temps longs m'avait conduit à fractionner les périodes de répétitions des spectacles. Il fallait alors organiser la structuration des développements technologiques et des expérimentations artistiques au fil de nombreuses étapes. Après une phase de recherche d'outil web et de préfiguration sur 2004, la plate-forme www.didascalie.net, basée sur la suite opensource tikiwiki, a été lancée en mars 2005 pour accompagner la réalisation d'*Escale#6* et constitue depuis un espace où j'invite mes collaborateurs à documenter et rendre accessible le plus de ressources possibles sur nos recherches, nos travaux artistiques et nos résultats. Cette formalisation incessante des processus créatifs en parallèle des temps de préparation et de répétitions représente travail conséquent et peu prisé des praticiens. Mais elle constitue une mémoire nécessaire à la prise de conscience des difficultés, facilite l'échange des points de vue et favorise l'inscription de la création dans le cadre d'une recherche.

---

7 Georges Gagneré, « Du développement "collectif" des outils du théâtre » in *Théâtres en Bretagne*, n°19, 1er semestre 2004, pp. 45-49.
8 *Escale#4* à Paris du 10 au 15 mai 2004. Cf. http://www.didascalie.net/escale4

Le fait le plus troublant était qu'il fallait prendre beaucoup de temps de préparation pour mettre en place un environnement immédiatement réactif pendant les répétitions. Je le nommais syndrome du « temps réel du temps réel »[9]. Avec le numérique et l'ordinateur, la notion d'interactivité rencontrait sur le plateau scénique l'inventivité de l'acteur et du metteur en scène. Habitués à travailler dans l'environnement matriciel d'un texte dramatique, ces deux collaborateurs multipliaient les propositions et les enrichissaient strate par strate au fil des répétitions. Avec l'élargissement du processus d'écriture scénique à des images, des sons, des capteurs, l'équipe artistique était confrontée à une dramaturgie en cours de fabrication, sollicitant incessamment les auteurs des matériaux scéniques : adaptation des textes, création des images et des sons. Face aux logiciels traitant en temps réel les réactions des comédiens et des régisseurs, il aurait fallu déployer une méthodologie qui rende possible, en temps réel, leur propre modification.

A partir du moment où Mirage a été stabilisé, notamment à l'occasion des *Escales #5* et *#6*, les enjeux d'utilisation se sont progressivement déplacés vers l'écriture de la dynamisation des effets. Nous réfléchissions beaucoup à l'époque sur la notion de « surface de contrôle » pilotant des moteurs vidéo ou audio, et qui aurait permis d'écrire la conduite des effets en dialogue étroit avec le metteur en scène. C'est notamment à l'occasion d'un travail sur des textes de Perec et Calvino durant *Escale#7*[10] qu'Olivier Pfeiffer, régisseur audionumérique, avait développé le logiciel Peralvino[11]. Et la collaboration avec un éclairagiste islandais utilisant Max, m'avait permis, pour la première fois, d'expérimenter une interaction globale entre le jeu et les régies numériques son, image et lumière, dépassant l'étape précédente de juxtaposition à laquelle nous étions parvenu sur *La Pensée*.

## Formalisation des questionnements (2005-06)

A la suite des réflexions intenses et des rencontres fructueuses de 2004, et sur les conseils de François Giroudon, rencontré à l'occasion d'*Escale#4*, j'ai alors déposé en mars 2005 un projet de groupe de travail qui fut retenu et financé par l'Association Française d'Informatique Musicale (AFIM). Francis Faber et Tom Mays de La Grande Fabrique de Dieppe, Pascal Baltazar, compositeur alors en résidence au GMEA d'Albi, François Weber enseignant à l'Institut Supérieur des Technique du Spectacle d'Avignon, mes collaborateurs et moi-même officialisaient une réflexion collective autour des outils logiciels traitant le son ou la vidéo et des « philosophies » d'utilisation de ces outils dans le cadre du spectacle vivant. En parallèle, j'explorais la champ académique de la recherche universitaire, et je fis la rencontre de Christian Jacquemin, enseignant-chercheur au LIMSI-CNRS, que j'interrogeais sur les meilleures méthodes de développement des moteurs video pour répondre aux enjeux d'écriture scénique interactive. Spécialiste d'informatique graphique, Christian Jacquemin venait de fabriquer pour ses recherches personnelles un environnement de composition en 3D, appelé Virchor, et qu'il souhaitait partager avec le monde du spectacle vivant.

---

9   Téléchargeable sur www.didascalie.net/tiki-download_file.php?fileId=443
10  *Escale#7*, du 4 au 16 avril 2005 à La Filature, Scène Nationale de Mulhouse. Cf. www.didascalie.net/escale7
11  Ce nom s'inspire de celui des deux auteurs : www.didascalie.net/Peralvino_zone

À l'occasion de mon spectacle suivant, *La Pluralité des Mondes*[12], conçu à partir d'un recueil du poète oulipien Jacques Roubaud, je proposais à Pedro Soler de travailler avec Christian Jacquemin sur la profondeur 3D de l'espace visuel et j'entamais une collaboration avec Tom Mays en vue d'approfondir l'interaction entre jeu et univers sonore. Cela eut un double effet sur notre environnement numérique de création. Ayant initialement prévu d'utiliser son logiciel Peralvino, Olivier Pfeiffer trouva cependant dans Tape, l'environnement de composition de Tom Mays, un outil lui permettant de poursuivre la structuration de la régie numérique tout en étant au plus près du travail artistique du compositeur. En réussissant à s'adapter à un logiciel moteur adéquat, le régisseur audio illustrait, selon moi, la pertinence d'une dissociation entre l'utilisation par une écriture scénique d'un environnement numérique complexe, et la fabrication de ce même environnement. Pour un metteur en scène, en effet, le régisseur est un collaborateur très précieux qui garantit la relation à la matérialité du plateau, qu'elle soit mécanique, électrique, électronique ou numérique. Cet état de fait provoqua de longues discussions au sein du groupe AFIM sur la frontière mouvante et difficilement définissable entre le régisseur et le créateur de média dans le spectacle vivant. Le régisseur participe étroitement au processus, conduit par le directeur artistique, de coordination des divers éléments scéniques et d'harmonisation des multiples énergies créatrices. L'introduction du paradigme numérique et de son corollaire d'interaction généralisée démultipliait le champ du possible tout en le complexifiant. *La Pluralité des Mondes* en offrait un exemple dans la circulation de la vidéo (Mirage), de la 3D (Virchor) et du son (Tape) en interaction avec Christophe Caustier, l'unique interprète. Les trois moteurs logiciels étaient pilotés par une conduite extérieure organisant les paramètres et imbriquant les interactions (cf. figure 3 c). C'est ainsi que Christophe Caustier jonglait avec un mur d'eau virtuel vers la fin du spectacle et contrôlait simultanément l'environnement sonore (cf. figure 3 a et b). Mélangeant 3D et vidéo, la scénographie devenait encore plus malléable que sur *La Pensée*, ce qui induisait de nouvelles règles d'écriture de la réalité scénique augmentée[13].

Au fil de l'année 2006, les réflexions du groupe AFIM s'intensifièrent et débouchèrent sur la volonté de faire converger les environnements numériques de création propres à chacun. La proximité des problématiques renforçait l'intuition d'une possible mutualisation des outils, préalable à une amélioration collective des difficultés rencontrées dans nos développements informatiques incessants. En parallèle, le projet Jamoma[14] avait démarré quelques mois auparavant et ses objectifs affichés de favoriser l'interopérabilité de patchs Max étaient suffisamment proches de ceux du groupe AFIM pour qu'un rapprochement s'opère et qu'émerge l'idée de prototyper un Environnement Versatile Expérimental[15] (EVE) de création commun, à partir de Jamoma. À la question initialement posée de savoir comment améliorer les outils numériques du spectacle vivant, Christian Jacquemin suggéra de fédérer des partenaires et de déposer une proposition dans le cadre des appels à projet du réseau pour la Recherche et l'Innovation en Audiovisuel et Multimédia de l'Agence Nationale de la Recherche. Le groupe AFIM formait un noyau naturel dont le rapprochement avec des laboratoires scientifiques pourrait consolider les approches recherche-création de chacun des partenaires. C'est ainsi que le projet Virage[16] vit le jour et fut ultérieurement sélectionné par l'ANR pour démarrer en décembre 2007, pour une durée de deux ans.

---

12  Création les 1 et 2 décembre 2005 à La Filature, Scène Nationale de Mulhouse. Cf. www.didascalie.net/pluralite
13  Christian Jacquemin, Georges Gagneré, « Image de synthèse temps réel pour la performance augmentée dans le spectacle vivant : Une interface de conception et de contrôle à base de calques physiques » in *Interfaces numériques*, sous la direction de Imad Saleh et Djeff Regottas, Hermès-Lavoisier, juin 2007, pp. 135-152
14  www.jamoma.org/
15  Pascal Baltazar, Georges Gagneré, « Outils et pratique du sonore dans le spectacle vivant » in *Actes des Journées d'informatiques Musicales* (JIM07), avril 2007, Lyon, pp. 153-162.
16  www.virage-platform.org/

# Une recherche stimulant la création (2007)

Dans la foulée d'*Escale#7 et La Pluralité des Mondes*, mon désir de poursuivre l'exploration de l'univers oulipien se concrétisa avec la mise en chantier d'*Espaces Indicibles*, d'après *Espèces d'espaces* de Georges Perec. Dès les premières répétitions à l'automne 2006, nous avons alors essayé d'utiliser Jamoma pour construire un moteur vidéo en suivant les spécifications d'EVE. Mais Jamoma était malheureusement encore trop jeune et fragile pour réaliser un moteur fiable en production. Travaillant avec Tape, le moteur audio stable de Tom Mays, je sollicitais ce dernier pour réfléchir à une nouvelle solution vidéo à partir des principes d'EVE et améliorant Mirage. Avec l'aide du compositeur Pascal Baltazar pour le développement, en dialogue avec Renaud Rubiano, plasticien et régisseur vidéo, le logiciel Movie, basé sur Max-Jitter, fut créé et utilisé en parallèle de Virchor sur le spectacle.

Parallèlement au dépôt du projet Virage en mars 2007, une première version d'*Espaces Indicibles* était créée en mai 2007 à Mulhouse, puis finalisé à l'automne 2007 dans le cadre du festival Musica de Strasbourg[17]. Cela nous procura deux étapes de répétition pour consolider Movie, et affiner les interactions entre les interprètes et l'environnement intermedia (cf. figure 4 c). J'avais demandé à la danseuse Mercé de Rande de se joindre à Christophe Caustier pour jouer dans le spectacle. Le comédien manipulait les sons et les images avec des capteurs embarqués et la danseuse interagissait par un tracking caméra. Dans les premiers moments du spectacle, Mercé de Rande introduisait physiquement l'espace en pointillé. Le régisseur déclenchait chaque tiret en fonction de ses mouvements, et la progression des déclenchements pilotait un patch générant progressivement l'environnement sonore. Cette écriture pas à pas de l'environnement scénique conférait à la fois une liberté de jeu à la danseuse, et l'amenait à véritablement s'appuyer sur ce qui l'entourait. Dans la séquence suivante, Christophe Caustier nous faisait voyager mentalement sur le texte de Perec, « La page » (« L'espace commence ainsi, avec seulement des mots, des signes tracés sur la page blanche »), et nourrissait une improvisation réalisée par la danseuse. Le régisseur captait à la volée des images de la danseuse et générait l'espace visuel avec la décomposition ralentie des moments de danse enregistrés en temps réel (cf. figure 4 a). La circulation d'énergie de l'acteur à la danseuse se cristallisait dans l'espace scénique. Ces deux séquences firent ultérieurement l'objet d'une description détaillée de la manipulation de l'environnement numérique afin d'alimenter le projet Virage en cas d'étude de « terrain ». Par ailleurs, le dispositif scénographique permettait de déployer une écriture de l'espace visuel donnant toute sa place à la 3D et constituait un aboutissement de l'intégration des comédiens dans une forme de réalité augmentée (cf. figure 4 b). En sept ans, les technologies numériques avaient rendu possibles les rêves de 2000.

Le projet Virage avait pour objectif « d'aboutir à une plate-forme autour de l'écriture du temps et de l'interaction dans la régie numérique de spectacle vivant »[18]. Je n'ai pas ici la place de le décrire en détail, d'autant plus qu'il a déjà fait l'objet de plusieurs publications et qu'il serait prématuré de tirer des conclusions puisqu'il se poursuit aujourd'hui sous la forme du projet OSSIA[19] jusqu'en 2015. On peut cependant dire que la réflexion collective dans le cadre de Virage a permis de faire émerger une question centrale, qui englobait les préoccupations des différents chercheurs et artistes associés. Comment écrire la dimension temporelle des scénarios interactifs composant les œuvres de chacun? Cette question recoupait une recherche déjà abordée par des laboratoires scientifiques et qui s'attache à élaborer un formalisme mathématique[20] pour représenter les différentes modalités de scénarisation temporelle, l'enjeu étant de pouvoir implémenter ce formalisme dans un ordinateur pour assister le créateur. Cette question temporelle constitue très probablement une des problématiques fondamentales d'un « laboratoire technologique de la mise en scène ».

---

17 Création les 11 et 12 mai 2007 à La Filature, Scène nationale de Mulhouse, puis reprises du 28/09 au 6/10/2007 au Théâtre National de Strasbourg. Cf. www.didascalie.net/espaces

18 Un bilan à mi-parcours est transcrit dans Pascal Baltazar et al., « Virage : Une réflexion pluridisciplinaire autour du temps dans la création numérique » in *Actes des Journées d'Informatique Musicale* (JIM09), 2009, Grenoble, pp. 151-160. Le compte-rendu final est dans Antoine Allombert et al., « Virage : Designing an interactive intermedia sequencer from users requirements and theoretical background » in *Proceedings of the International Computer Music Conference*, juin 2010, New York, pp. 470-473.

19 Théo De La Hogue, et al., « OSSIA : Open Scenario System for Interactive Application » in Actes des Journées d'informatiques Musicales (JIM14), mai 2014, Bourges, pp. 78-84.

## La fragilité motrice de la recherche-création

L'expérience que je viens d'évoquer est caractérisée par des allers-retours incessants entre l'expérimentation sur le plateau et le développement d'outils numériques de création. Se mettre à l'écoute des transformations nécessaires constitue un chantier à part entière dont il n'existe pas de méthode *a priori*. La recherche se construit au grès des rencontres individuelles artistiques et techniques, des échecs et des succès, des contraintes économiques. Elle se nourrit par une mise en réseau avec d'autres pratiques technologiques, artistiques et socio-culturelles. Les créations produites sont fragiles car elles reposent sur des prototypes et sont soumises à l'influence de trois dynamiques parfois contradictoires :

- favoriser l'émergence d'un espace d'écriture commun, cohérent et accessible à tous les collaborateurs artistiques et techniques des projets ;
- conduire les développements technologiques correspondant aux ambitions d'écriture interactive temps réel dans le contexte incontournable de la mutualisation
- préserver la liberté de création qui conduit inévitablement à une identité spécifique pour chaque œuvre.

Ces forces incitent à la convergence des recherches tout en laissant s'exprimer les singularités créatives. Elles produisent un équilibre perpétuellement déstabilisé, dont le risque vital est aussi le moteur de la création et de la recherche.

Dans le cadre de Virage, nous avions prévu de faire aboutir le chantier collectif de la maquette EVE, garantie de convergence et de mutualisation. Mais des retards légitimes dans le développement de Jamoma nous ont conduits une seconde fois à abandonner Jamoma, rappelant combien le phasage temporel de chantiers complexes est difficilement maîtrisable. Mes collaborateurs et moi-même avons dû nous résoudre à conserver les moteurs Tape et Movie, que nous avons fusionnés dans le moteur intégré Tapemovie[21] pour accéder aux outils produits par Virage, et notamment au séquenceur qui en constituait le résultat principal. Ce séquenceur matérialisait en effet le formalisme temporel auquel le projet avait abouti et devait permettre aux équipes artistiques de contrôler leurs différents moteurs. Malheureusement, le séquenceur Virage alourdissait considérablement l'utilisation de Tapemovie et n'était pas utilisable en création. Cette perte d'équilibre est bien entendu momentanée, car la recherche se poursuit aujourd'hui avec le projet OSSIA qui apportera certainement de nouvelles solutions. Il est possible aussi que l'émergence d'une problématique théorique doive oublier un moment les réalités empiriques pour se cristalliser et revenir ensuite se frotter au terrain. Cela promet de nouvelles explorations qui nous conduiront vers une appropriation encore plus intime du paradigme numérique.

Georges Gagneré, août 2014

---

20 Antoine Allombert, Gérard Assayag, Myriam Desainte-Catherine, « De boxes à Iscore : vers une écriture de l'interaction » in Actes des Journées d'Informatique Musicale (JIM08), mars 2008, Albi

21 Tom Mays, Renaud Rubiano, « Tapemovie : un environnement logiciel pour la création temps réel intermedia » in Actes des Journées d'Informatique Musicale (JIM10), mai 2010, Rennes, pp. 207-213. Cf. https://didascalie-net.github.io/tapemovie/

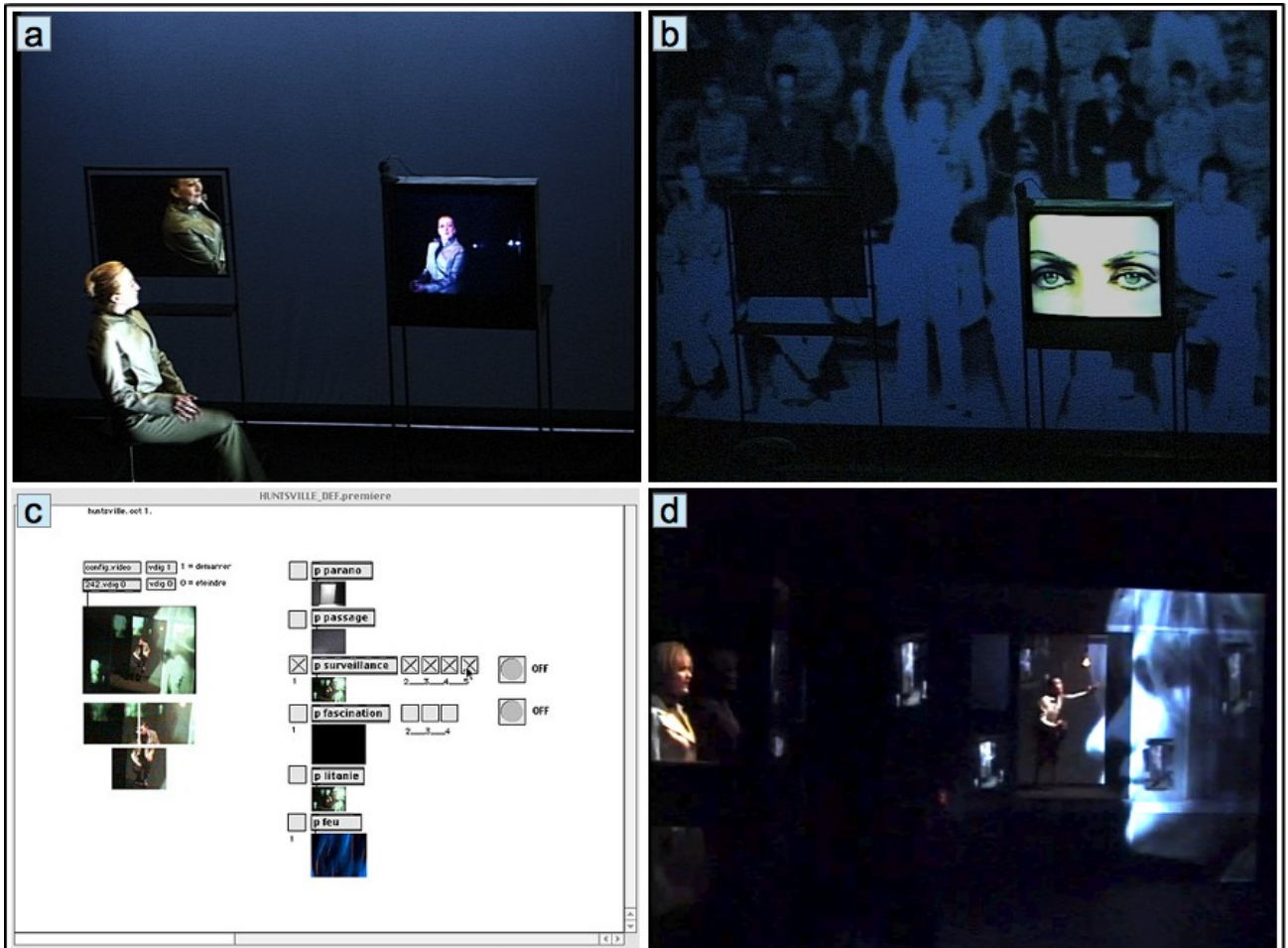

**Figure 1**. *H Manifeste[s]* (2000). Madame H (Isabelle Olive) se parle en direct dans le téléviseur (a). Elle joue dans le public face à son reflet en direct sur l'écran de rétroprojection (b).
*Huntsville, l'Ordre du Monde* (2001). Patch NATO de la partie numérique de la conduite vidéo (c), correspondant à l'incrustation en direct du visage de la Femme en Bleu (Delphine Raoult) dialoguant avec le Gardien (Eric Jakobiak), emprisonné dans sa cage de vidéosurveillance (d). Crédits : Georges Gagneré.

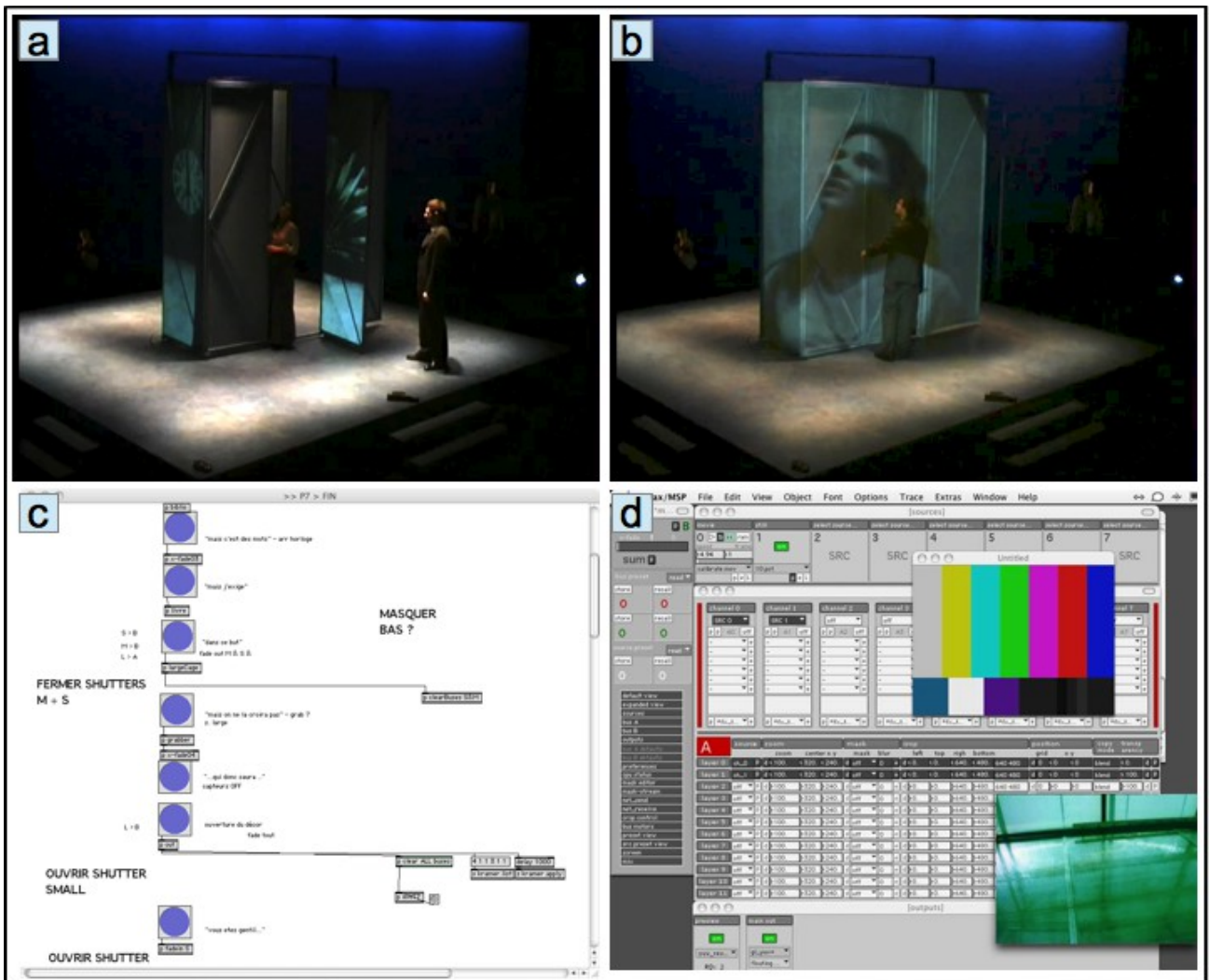

**Figure 2**. *La Pensée* (2003). Kerjentsev (Christophe Caustier) fantasme sur Tatiana (Margot Faure) (a), avant de l'emprisonner dans son image (b). Un extrait de la conduite numérique (c) pilotant le moteur vidéo Mirage (d). Crédits : Georges Gagneré.

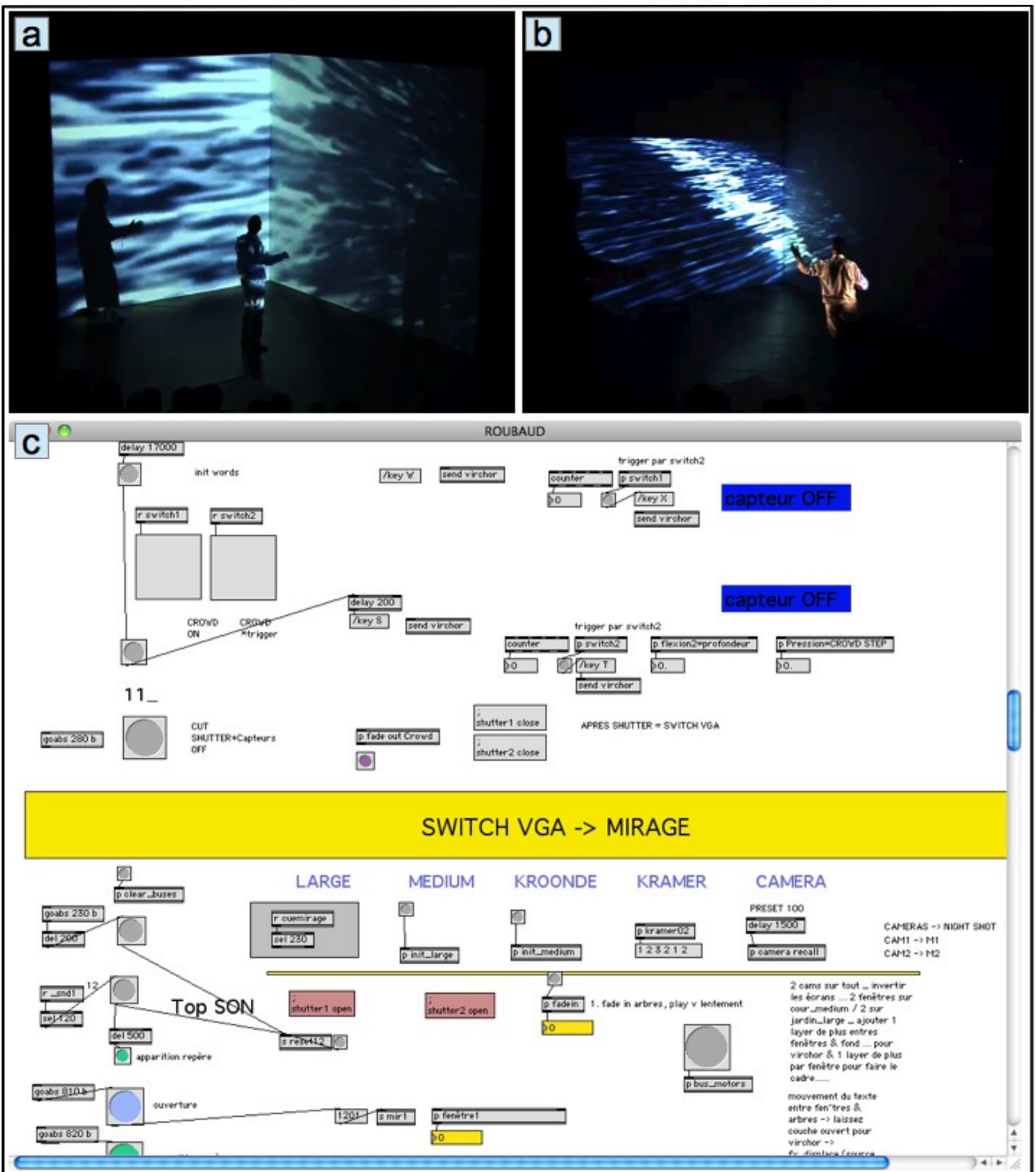

**Figure 3**. *La Pluralité des Mondes* (2005). Christophe Caustier manipule un mur d'eau avec des capteurs sans fil embarqués en zoomant (a) ou en l'orientant dans l'espace (b). Extrait de la conduite numérique pilotant l'ensemble de la vidéo et des images 3D (c). Crédits : Georges Gagneré.

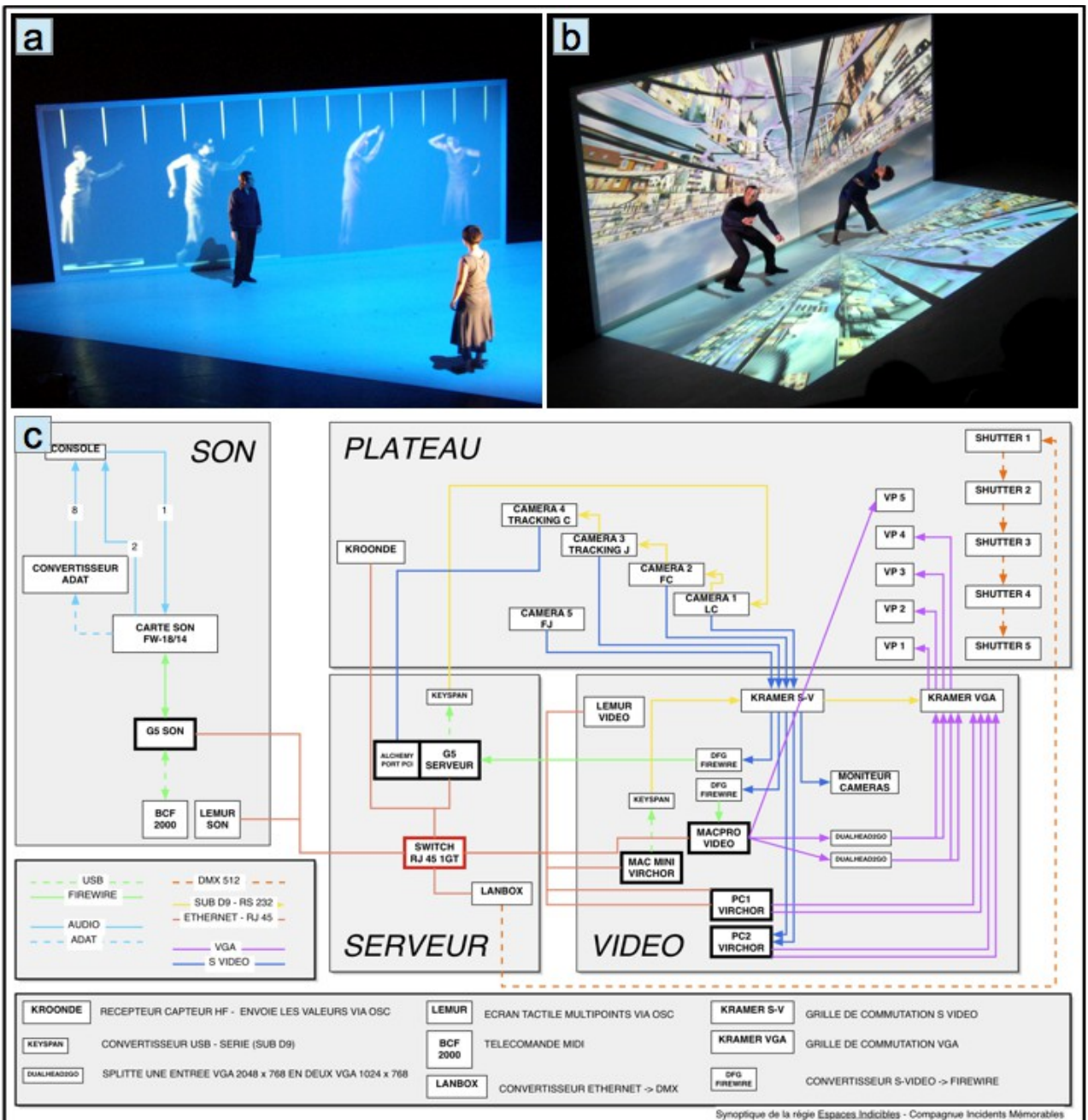

**Figure 4**. *Espaces Indicibles* (2007). Capture de quatre moments d'improvisation dansée (a) de Mercé de Rande. Christophe Caustier et la danseuse habitent les espaces 3D construits par Christian Jacquemin et le logiciel Virchor (b). Synoptique de la régie numérique (c). Crédits : Georges Gagneré.